# The Transformative Value of Liberating Mars

Jacob Haqq-Misra

Blue Marble Space Institute of Science
1200 Westlake Ave N Suite 1006**,** Seattle WA 98109
jacob@bmsis.org

**ABSTRACT**
Humanity has the knowledge to solve its problems but lacks the moral insight to implement these ideas on a global scale. New moral insight can be obtained through transformative experiences that allow us to examine and refine our underlying preferences, and the eventual landing of humans on Mars will be of tremendous transformative value. Before such an event, I propose that we *liberate* Mars from any controlling interests of Earth and allow Martian settlements to develop into a second independent instance of human civilization. Such a designation is consistent with the Outer Space Treaty and allows Mars to serve as a test bed and point of comparison by which to learn new information about the phenomenon of civilization. Rather than develop Mars through a series of government and corporate colonies, let us steer the future by liberating Mars and embracing the concept of planetary citizenship.

## EXTENDING OUR MORALITY

Technological solutions to humanity's major problems are known. Climate change can be addressed through a strategy of mitigation, adaptation, and (if needed) geoengineering in order to reduce our emissions and find cleaner sources of energy[1,2]. The "tragedy of the commons" that arises from an overpopulated planet can be alleviated through voluntary restrictions on breeding and changes in personal lifestyle[3]. Humanity can even survive long-term changes in the sun by first engaging in geoengineering[4] and eventually migrating to space settlements[5-7]. But if we already possess the knowledge to overcome these challenges, then why do such problems persist?

Ecologist Garrett Hardin noted that solving the population problem "requires a fundamental extension in morality"[3], and similar claims can be made about our failed efforts to address climate change, poverty, economic injustice, and other global challenges. One resolution to this dilemma is Plato's "philosopher king" or a similar "benevolent dictator" who enacts fair and consistent solutions to global problems. Barring such a non-reality, humans must find a way to develop new moral tools that will help to reduce the environmental impact of our civilization. There is no shortage of suggestions for how humanity should extend its morality, and the problem is often reduced to choosing among competitors. Contemporary political, religious, economic, or philosophical discourse is unlikely to develop solutions to these global problems, but it remains unclear as to how else new modes of thinking can be discovered.

**TRANSFORMING OUR PREFERENCES**
Certain experiences can lead us to examine our beliefs, ideas, or preferences in a way that provides deeper insight or novel understanding. A trust fund child who lives a life of luxury is whisked away by friends on a canoe trip where he finds greater joy in nature than any of his material wealth. A concert violinist is taken against her will to an improvisational rock concert, but her musical senses are so aroused that she abandons her classical career to pursue modern jazz. Experiences such as these have been described by Bryan Norton[8] and other philosophers[9,10] as sources of transformative value. Transformative experiences can challenge our core preferences, often at unexpected times, and force us to consider or conceptualize new perspectives. Transformative events can catalyze permanent epistemic change in individuals or communities that would have been impossible in the absence of such experiences[9].

Transformative experiences have guided the development of civilization and led humanity toward new ways of thinking. Transformative value often arises from experiences that challenge the boundaries between humans and environment as a result of ongoing exploration of the physical world[10]. Sources of transformative value throughout history include great composers, artists, performers, teachers, and leaders who inspired multiple generations and whose influence persists today. Scientific transformations through the Agricultural, Copernican, Darwinian, and Industrial Revolutions refined our physical knowledge, while images of our planet from space—such as the "Blue Marble" from Apollo 17 and the "Pale Blue Dot" from Voyager 1—encouraged the birth of environmentalism and Earth system science. Transformative experiences allow the unexpected realization of new ideas, but the impact will vary from person to person in ways that are impossible to predict. Even when an experience seems likely to have great transformative value (such as landing on the Moon, or sequencing the human genome, or discovering the Higgs boson), it can be difficult to determine exactly how it will affect our preferences.

Humanity can steer the future by searching for sources of transformative value that will help us to achieve the needed "extension in morality". But how do we identify desirable or beneficial sources of transformative value from those that might be undesirable or harmful? The unpredictable character of transformative experiences makes this question difficult to answer[8,9]. Perhaps we should avoid indoctrinating experiences that seek to conform or restrict our preferences. Perhaps we should seek experiences that confront our boundaries between environment and self[10]. Transformative experiences on a local scale may help to improve individual lives, but can we strive toward a transformative goal that will give our entire civilization a needed change in perspective? I suggest that Mars holds the answer to this question.

**MARS AS THE NEXT FRONTIER**
Mars has long been a source of wonder for humans, and the first step of a human on Martian soil will carry incalculable transformative value to the people of Earth. Several private corporations have declared their intent to send humans to Mars within the next several decades, including SpaceX, MarsOne, and the Inspiration Mars Foundation. MarsOne, in particular, seeks to establish a permanent colony on Mars by 2023 and hopes to

"finance this mission by creating the biggest media event ever"[11]. Whether or not they succeed, such endeavors indicate increasing public interest in space colonization and suggest that private corporations, rather than government space agencies, may be the first to reach Mars.

Any entity that establishes a permanent colony on Mars is, in effect, laying claim to the land on another planet. Such claims to ownership of a celestial body are explicitly forbidden by the Outer Space Treaty of 1967, which states that, "Outer space, including the Moon and other celestial bodies, is not subject to national appropriation by claim of sovereignty, by means of use or occupation, or by any other means". The treaty establishes space as "the province of all mankind" that should be "free for exploration and use by all states without discrimination of any kind, on a basis of equality" with "free access to all areas of celestial bodies", indicating that no nation can claim exclusive access to Mars or any other celestial bodies[12,13]. This creates a similar provision as in the Antarctic Treaty System that prohibits national claims to land; however, the language of the Outer Space Treaty is sufficiently vague that it does not necessarily discuss implications for individuals or corporations who venture into space.

Further problems arise if we imagine the continued exploration of Mars by the various nations of the world: Mars, like Earth, contains finite resources that cannot be shared equally by all, so is it equitable to allow certain nations to benefit from Mars exploration more than others? If, for example, most of Mars' surface area were quickly colonized by US corporate explorers, then there might be little room left for European, Chinese, or Indian colonists who also may want to establish a presence on Mars. Existing treaties prohibit claims to ownership and are unclear about the use of space resources, so new policies are needed to govern land use on Mars over the coming centuries.

Finding a balance between human exploration and national interests is difficult and may require extensions or modifications to existing treaties. One solution is to adopt a model based upon the Antarctic Treaty System or the UN Convention on the Law of the Seas that would allow a stepwise approach toward balancing scientific and commercial interests[12]. Another option would seek to stimulate commercial interest in space by developing policies that allow for land ownership while still maintaining the spirit of the Outer Space Treaty[13]. But rather than give Mars over to the nations and corporations of Earth, I have a different idea: I propose that we *liberate* Mars.

**LIBERATING MARS**
How can the settlement of Mars transform our preferences and allow us to solve our global problems? If Mars is approached as an extension of Earth civilization, to be divided by the nations and plundered by industry, then no such transformation will occur. I therefore suggest that the goal of colonization should be not to extend our present civilization into new terrain but instead to create an independently functioning human civilization on Mars. If we wish to solve the political and economic problems that still plague us, and if we wish to discover new options for our development, then Mars provides an opportunity to create a second experiment in civilization.

My suggestion is to allow humans to permanently settle on Mars for the purpose of developing a self-sufficient Martian civilization. Although the Outer Space Treaty already prohibits any claims to national sovereignty, I suggest the following provisions also apply to the settlement of Mars:

1. Humans who leave Earth to permanently settle on Mars relinquish their planetary citizenship as *Earthlings* and claim a planetary citizenship as *Martians*. This includes giving up any national or local citizenships and affiliations. Humans living as Martians cannot represent the interests of any group on Earth and cannot acquire wealth on Earth.

2. Governments, corporations, and individuals of Earth cannot engage in commerce with Mars and cannot interfere with the political, cultural, economic, or social development of Martian civilization.

3. Scientific exploration may continue as long as it does not interfere with the development of civilization on Mars. Sharing of research and information between Mars and Earth is permitted only to pursue mutual scientific or educational goals.

4. The use of land on Mars will be determined exclusively by the citizens of Mars. No Earthlings may own or otherwise lay claim to land on Mars.

5. Any technology, resources, or other objects brought from Earth to Mars become permanent fixtures of the Martian civilization. Earthlings may not make any demands for resources on Mars.

These provisions would make the settlement of Mars contrast starkly with historical patterns of colonization on Earth. By *liberating* Mars according to this set of provisions, the red planet becomes accessible to humanity for the development of a new civilization but barred from ever being controlled by existing groups on Earth. This independent parallel development of civilization on Mars will provide a test bed for new ideas that could lead to unforeseen epistemic transformations of our values and preferences[8-10].

    As an example, suppose that a crew from the SpaceX corporation is the first to establish a settlement on Mars. According to the provisions of a liberated Mars, these individuals become citizens of planet Mars and abandon their political and economic ties with Earth. The crew is free to use Mars as they please in order to build the foundations of their settlement, and no entity on Earth—including SpaceX—can exert control over their actions. Any equipment brought with the crew permanently becomes Martian property; gifts or aid provided by Earth can be accepted but only without remuneration, and trade is strictly prohibited. The Martian settlers will have complete control over their new planetary home until another crew arrives—suppose this time from China. How the Chinese crew and SpaceX crew decide to use the land and resources of Mars is a dispute to be settled among themselves alone, without any interference from nations or corporations of Earth. Other settlers will eventually follow suit as well, which will require the citizens of Mars to develop their own laws and systems of conflict resolution. Eventually the scattered Martian settlements will slowly transform into an autonomous Martian civilization with governments, economies, and cultures of its own design. Ideally the settlers will draw on knowledge from human history to avoid repeating mistakes of the past, and the absence of control from Earth will allow the citizens of Mars to develop practical solutions to their problems.

    The decision to liberate Mars may be unpopular among governments and corporations today, particularly those who invest heavily in space exploration technologies, but a liberated

Mars would provide a new frontier for homesteading and a promise of a fresh start. This would change many existing motivations for space exploration, such as commercial or defense interests, but would open the door to space settlement for the purpose of developing a second human civilization on Mars. I cannot predict the outcome of this experiment and cannot anticipate how our preferences may be transformed; however, I can guarantee that if Mars colonization follows the patterns of history, then the ability of Mars to transform our morality will be lost forever.

There is no rush to settle Mars, but we should decide to liberate Mars before the first humans explorers arrive. Humans are nomadic by heritage, and we have not fully internalized the realization that our planet is a finite resource. But if we can muster the courage to place Mars firmly off limits to the interests of Earth, then we will begin to understand what it means to be planetary citizens.

**FROM HERE TO THERE**

Humanity requires an "extension in morality", and the liberated settlement of Mars provides an opportunity for approaching this transformation. If Mars is free from the interests of Earth, then any humans who are willing to relocate to the red planet will help to shape the future of Martian civilization. What forms of government, statehood, currency, spirituality, and culture will develop on Mars? If we liberate Mars from controlling interests on Earth, then we will discover new ideas, unexpected solutions, and keen insight about civilization itself that would otherwise be impossible.

Getting the first humans to Mars is the initial challenge, and establishing a self-sufficient settlement will be a daunting task to say the last. Critics of this idea may argue that a Martian colony, supported by nations or corporations of Earth, will have the greatest chance of success, and that liberation should be considered only after a series of successful colonies. I maintain that this will be too late: once the interests and ideals of Earth take root on Mars, they will be very difficult to supplant. A second instance of civilization is a better use of Mars than an extension of our resource base, if only because it holds the potential to transform our core preferences and conceptualize new civilizational and environmental values. Better to liberate Mars now and see who will heed its call.

Our current approach to space exploration has followed patterns of history so far, and we must make a conscious effort to find new ways of thinking to avoid repeating past mistakes. Mars is on our horizon, and the way we decide to use Mars will permanently transform the future of humanity. As we fix our eyes on the red planet, is our goal to extend our old ways into space to see how long they can last, or should we try bold new experiments in civilization to find better ideas? I choose the latter: Mars should belong to the Martians, *especially if they are human*.

**ACKNOWLEDGMENTS**

This paper was originally written for the 2014 Foundational Questions Institute (FQXi) Essay Contest addressing "How Should Humanity Steer the Future?". All opinions are those of the author alone. No funding was used for the development of this concept.